\documentclass[aps,prd,eqsecnum,preprint,tightenlines,nofootinbib,showpacs]{revtex4-1}
\usepackage{graphicx,amsmath,latexsym}

\newcommand{\bea}{\begin{eqnarray}}
\newcommand{\eea}{\end{eqnarray}}
\newcommand{\be}{\begin{equation}}
\newcommand{\ee}{\end{equation}}
\newcommand{\ub}[1]{\underline{#1}}
\newcommand{\ob}[1]{\overline{#1}}
\newcommand{\Pminus}{{\cal P}^-}

\begin{document}

\title{Light-front holography \\
and the light-front coupled-cluster method%
\footnote{Based on a talk contributed to the
Lightcone 2013 workshop, Skiathos, Greece, 
May 20-24, 2013.}
}

\author{J.R. Hiller}
\affiliation{Department of Physics \\
University of Minnesota-Duluth \\
Duluth, Minnesota 55812}

\date{\today}

\begin{abstract}
We summarize the light-front coupled-cluster (LFCC) method for the solution
of field-theoretic bound-state eigenvalue problems and indicate the
connection with light-front holographic QCD.  This includes a sample
application of the LFCC method and leads to a relativistic quark model
for mesons that adds longitudinal dynamics to the usual transverse
light-front holographic Schr\"odinger equation.
\end{abstract}

\maketitle

\section{Introduction} \label{sec:intro}

In order to compute hadronic light-front wave functions, we need
a method by which the light-front QCD Hamiltonian eigenvalue
problem can be solved nonperturbatively.  The standard
approach is to expand the eigenstate in a truncated Fock basis,
with the wave functions as the expansion coefficients, and solve
the resulting integral equations for these wave functions.
The light-front coupled-cluster (LFCC) method~\cite{LFCC} follows this
path, except that the Fock basis is not truncated; instead, the
wave functions for higher Fock states are restricted to being
determined from the wave functions for the lowest states through
functions that satisfy nonlinear integral equations.  In the
lowest Fock sector, designated the valence sector, there 
is an eigenvalue problem for an
effective LFCC Hamiltonian that approximates the effects
of higher Fock states.  It is this restricted eigenvalue
problem that is related to the light-front holographic
eigenvalue problem~\cite{holographicQCD}, which is usually presented
in the form of a transverse light-front Schr\"odinger equation
for massless quarks.  We extend light-front holography to
include a longitudinal equation and masses for quarks~\cite{LongWF}.

\begin{figure}[hb]
\begin{center}
\includegraphics[width=10cm]{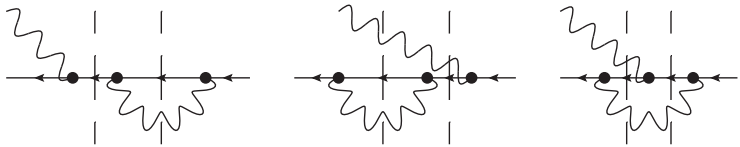}
\end{center}
\caption{\label{fig:WardID}One-loop diagrams contributing to the Ward
identity.  When only one intermediate photon is allowed, only one
diagram survives and the Ward identity fails.}
\end{figure}

The truncation of the Fock basis should be avoided, because
it causes uncanceled divergences.
The analog in Feynman perturbation theory is to 
separate diagrams into time-ordered diagrams and
discard time orderings that include intermediate states
with more particles than some finite limit.  This
destroys covariance, disrupts regularization, and 
induces spectator dependence for subdiagrams.
For example, the Ward identity of gauge theories is
destroyed by truncation, as illustrated in Fig.~\ref{fig:WardID}.  
In the nonperturbative case, this happens
not just to some finite order in the coupling but to all
orders.  The LFCC method is designed to avoid this sort
of complication.

As its name implies, the LFCC method is for light-front quantized
Hamiltonians.  We define light-front coordinates~\cite{Dirac} as the light-front
time $x^+=t+z$ and space $\ub{x}=\left(x^-\equiv t-z,\vec{x}_\perp=(x,y)\right)$.
The corresponding light-front energy and momentum are
$p^-=E-p_z$ and $\ub{p}=\left(p^+\equiv E+p_z,\vec{p}_\perp=(p_x,p_y)\right)$.
These imply that the mass-shell condition $p^2=m^2$ becomes
$p^-=\frac{m^2+p_\perp^2}{p^+}$, and the mass eigenvalue problem is
\be
\Pminus|\ub{P}\rangle=\frac{M^2+P_\perp^2}{P^+}|\ub{P}\rangle, \;\;\;\;
\underline{\cal P}|\ub{P}\rangle=\underline{P}|\ub{P}\rangle.
\ee

Light-front coordinates have several advantages~\cite{DLCQreviews}.
There are no spurious vacuum contributions to eigenstates, because
$p^+>0$ for all particles, which prevents particle production
from the vacuum without violation of momentum conservation.\footnote{This
can, of course, be defeated by zero modes, which can be important for
analysis of properties usually associated with vacuum 
structure~\protect\cite{LFCCzeromodes}.}
This permits well-defined Fock-state wave functions with no spurious
vacuum contributions.  Also, there is a boost-invariant separation
of internal and external momenta, so that wave functions can depend
on internal momenta only, which are usually taken to be the
longitudinal momentum fractions $x_i\equiv p_i^+/P^+$
and relative transverse momenta 
$\vec k_{i\perp}\equiv \vec p_{i\perp}-x_i\vec P_\perp$.

\section{Light-front coupled-cluster method}  \label{sec:LFCC}

The LFCC method solves the Hamiltonian eigenvalue problem by first writing
the eigenstate as $|\psi\rangle=\sqrt{Z}e^T|\phi\rangle$, where
$|\phi\rangle$ is the valence state, with normalization
$\langle\phi'|\phi\rangle=\delta(\ub{P}'-\ub{P})$, and
$T$ is an operator that increases particle number.  The overall
normalization is set by $Z$, such that
$\langle\psi'|\psi\rangle=\delta(\ub{P}'-\ub{P})$.
$T$ conserves all quantum numbers, including $J_z$, light-front momentum $\ub{P}$, 
and charge.  Because $p^+$ is positive, $T$ must include annihilation,
and powers of $T$ include contractions.
The LFCC effective Hamiltonian is constructed as 
$\ob{{\cal P}-}=e^{-T} \Pminus e^T$.  Then, with
$P_v$ the projection onto the valence Fock sector, we have the
coupled system
\be
P_v\ob{\Pminus}|\phi\rangle=\frac{M^2+P_\perp^2}{P^+}|\phi\rangle, \;\;\;\;
(1-P_v)\ob{\Pminus}|\phi\rangle=0.
\ee

Formulated in this fashion, the eigenvalue problem is still exact but also 
still infinite in size, because there are, in general, infinitely many 
terms in $T$.  To have a finite problem, but without truncation of the
Fock basis, we truncate $T$ and $1-P_v$.  The simplest truncation of
$T$ is to include single-particle emission, such as a gluon from a
quark; the corresponding truncation of $1-P_v$ would be to project
onto Fock states with one more gluon than the valence state.  The
truncation of $1-P_v$ then provides just enough equations to solve
for the emission vertex function contained in $T$.  The truncations
can be systematically relaxed, by expanding the number of particles
created by $T$ and the range of Fock states used for projections.

The mathematics of the LFCC method have their origin in the 
nonrelativistic many-body coupled-cluster method~\cite{CCorigin}, developed
in nuclear physics and quantum chemistry~\cite{CCreviews}.
It was first applied to the many-electron problem in molecules by
\v{C}i\v{z}ek~\cite{Cizek}.  The Hamiltonian eigenstate is formed as
$e^T|\phi\rangle$, where $|\phi\rangle$ is a product of single-particle states
and where terms in $T$ annihilate states in $|\phi\rangle$ and create excited states,
to build in correlations.  The operator $T$ is then truncated
at some number of excitations; however, the number of particles
does not change.  There are also some applications to quantum field theory
in equal-time quantization~\cite{CC-QFT}.

Once the LFCC eigenvalue problem has been solved, the solution
can be used to compute observables, such as form factors.
For a dressed fermion, the Dirac and Pauli form factors
can be computed from a matrix element of the current 
$J^+=\ob{\psi}\gamma^+\psi$, which 
couples to a photon of momentum $q$.
The matrix element is generally~\cite{BrodskyDrell}
\be
\langle\psi^\sigma(\ub{P}+\ub{q})|16\pi^3J^+(0)|\psi^\pm(\ub{P})\rangle
=2\delta_{\sigma\pm}F_1(q^2)\pm\frac{q^1\pm iq^2}{M}\delta_{\sigma\mp}F_2(q^2),
\ee
with $F_1$ and $F_2$ the Dirac and Pauli form factors.
Thus, we need to be able to compute matrix elements.

As an example of how a matrix element can be computed, we consider
the expectation value for an operator $\hat{O}$, which in the LFCC method
would be expressed as
\be 
\langle\hat O\rangle=\frac{\langle\phi| e^{T^\dagger}\hat O e^T|\phi\rangle}
                      {\langle\phi| e^{T^\dagger} e^T|\phi\rangle}.
\ee
Direct computation would require an infinite sum over the untruncated Fock basis.
Instead, we define $\ob{O}=e^{-T}\hat O e^T$ and 
$\langle\tilde\psi|=\langle\phi|\frac{e^{T^\dagger}e^T}
      {\langle\phi|e^{T^\dagger} e^T|\phi\rangle}$,
so that $\langle\hat O\rangle=\langle\tilde\psi|\ob{O}|\phi\rangle$ and
$\langle\tilde\psi'|\phi\rangle
=\langle\phi'|\frac{e^{T^\dagger}e^T}{\langle\phi| e^{T^\dagger} e^T|\phi\rangle}|\phi\rangle
=\delta(\ub{P}'-\ub{P})$.
The effective operator $\ob{O}$ is computed from the  Baker--Hausdorff expansion,
$\ob{O}=\hat O + [\hat O,T]+\frac12[[\hat O,T],T]+\cdots$.
The bra $\langle\tilde\psi|$ is a left eigenvector of $\ob{\Pminus}$, because the
following holds:
\be
\langle\tilde\psi|\ob{\Pminus}
=\langle\phi|\frac{e^{T^\dagger}\Pminus e^T}{\langle\phi| e^{T^\dagger} e^T|\phi\rangle}
=\langle\phi|\ob{\Pminus}^\dagger \frac{e^{T^\dagger}e^T}
                            {\langle\phi| e^{T^\dagger} e^T|\phi\rangle}
=\frac{M^2+P_\perp^2}{P^+}\langle\tilde\psi|.
\ee
With this technique, the Dirac form factor is approximated by the matrix element
\be \label{eq:DiracFF}
F_1(q^2)=8\pi^3\langle\widetilde\psi^\pm(\ub{P}+\ub{q})|\ob{J^+(0)}|\phi^\pm(\ub{P})\rangle,
\ee
with $\ob{J^+(0)}=J^+(0)+[J^+(0),T]+\cdots$.

\section{Sample LFCC application} \label{sec:model}

\begin{figure}
\begin{center}
\includegraphics[width=12cm]{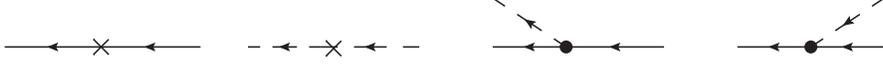}
\end{center}
\caption{\label{fig:Pminus}
Graphical representation of the light-front Hamiltonian $\Pminus$
for the soluble model discussed in Sec.~\ref{sec:model}.  A
cross represents a kinetic energy contribution.  Lines on the right
represent annihilation operators; lines on left, creation operators.
Solid lines are for the fermion, and dashed lines, for the boson.}
\end{figure}

As an example of the use of the LFCC method~\cite{LFCC}, we consider
a soluble model, a light-front analog~\cite{model} of the 
Greenberg--Schweber model~\cite{Greenberg}.  In this model,
a heavy fermionic source emits and absorbs bosons without
changing its spin, and we solve for the fermionic eigenstate
dressed by a cloud of bosons.  A graphical representation 
of the light-front Hamiltonian $\Pminus$ is given in Fig.~\ref{fig:Pminus}.

The model is not fully covariant; in
particular, states are all limited to having a fixed total
transverse momentum.  This hides some of the power of the
LFCC method, but is sufficient to show how the method can
be applied.  Details can be found in Ref.~\cite{LFCC}.

\begin{figure}
\begin{center}
\begin{tabular}{c}
\includegraphics[height=1cm]{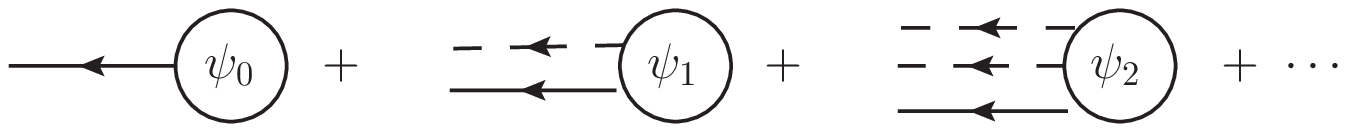} \\ (a) \\
\includegraphics[height=4.5cm]{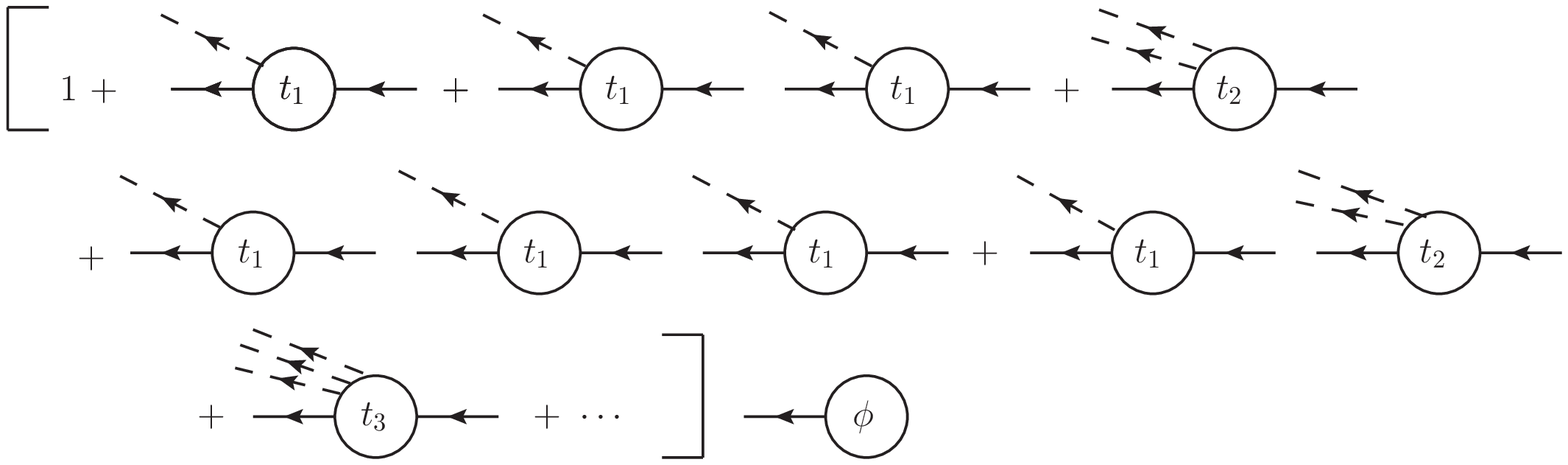} \\ (b)
\end{tabular}
\end{center}
\caption{\label{fig:psi}
The model eigenstate as (a) Fock-state expansion
and (b) LFCC construction. Bosons are represented
by dashed lines, and the fermion by a solid line.
The $\psi_n$ are the
Fock-state wave functions; the $t_n$ are the functions
that determine the structure of the $T$ operator,
by setting the momentum distribution of the $n$
created bosons.  The bare fermion state is
represented by $\psi_0$ in (a) and by $\phi$ in (b).  
The line to the right of each
$t_n$ bubble represents a fermion annihilation operator
that is contracted with a creation operator in
the bare fermion state or the next $t_n$ vertex.}
\end{figure}

Here we compare the LFCC method with a traditional
truncated-Fock-space approach.  The Fock-state expansion
of the eigenstate is represented in Fig.~\ref{fig:psi}(a).
For the LFCC method, we take the valence state to
be the bare fermion.  The
resulting LFCC form of the eigenstate is represented
in Fig.~\ref{fig:psi}(b).

\begin{figure}
\begin{center}
\begin{tabular}{c}
\includegraphics[height=3.5cm]{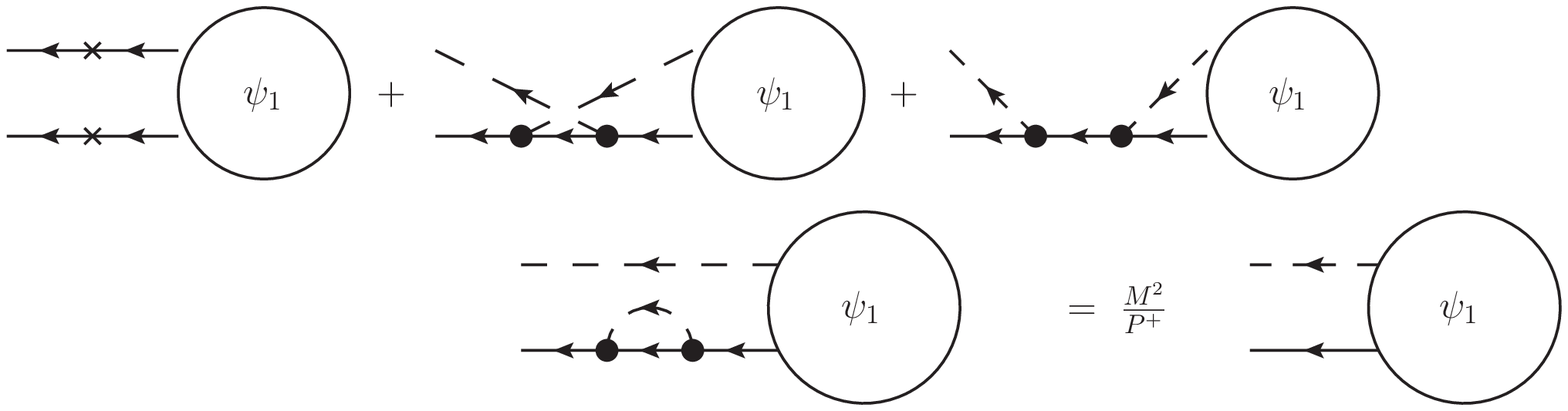} \\ (a) \\
\includegraphics[height=1.5cm]{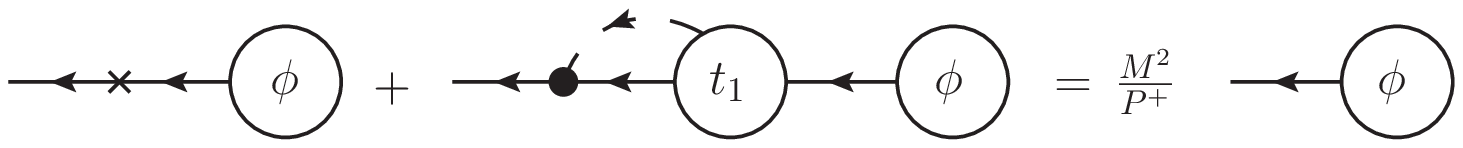} \\ (b) \\
\includegraphics[height=3.5cm]{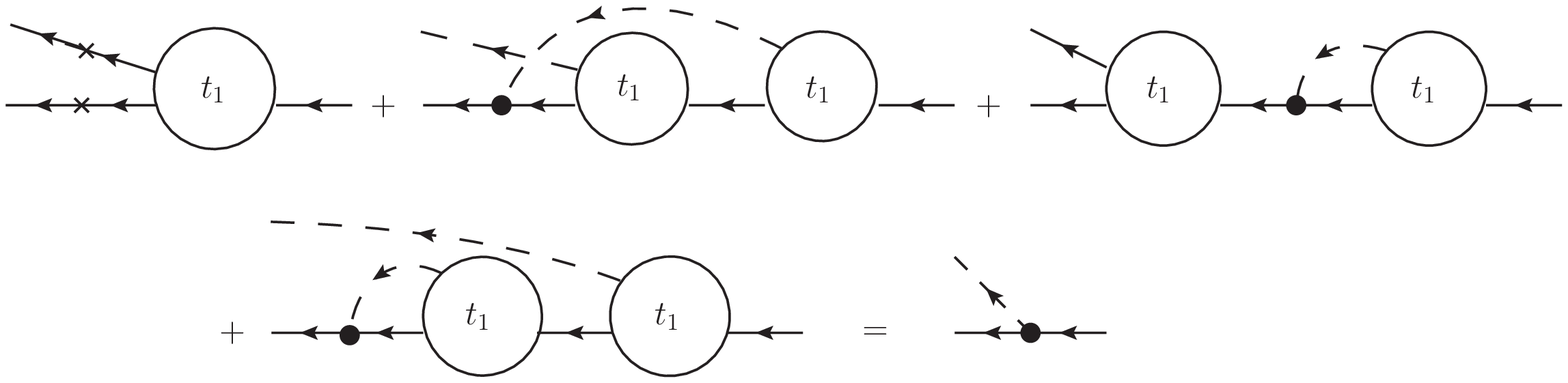} \\ (c)
\end{tabular}
\end{center}
\caption{\label{fig:eqn}
Graphical representations of the integral equations
for (a) the Fock-state wave function $\psi_1$,
(b) the valence state $|\phi\rangle$,
and (c) the function $t_1$ that determines the
truncated $T$ operator.  The dots represent the boson
emission/absorption vertex of the model Hamiltonian.}
\end{figure}

We truncate the $T$ operator to include only single
boson emission, which corresponds to the terms with
$t_1$ in Fig.~\ref{fig:psi}(b).  A comparable truncation
of the Fock-space expansion is to limit the number
of bosons to no more than two.  The resulting integral
equations are represented in Fig.~\ref{fig:eqn}.
In the Fock-space truncation case, the self-energy
contribution is spectator dependent, with an energy
denominator that includes the second boson in flight
as well as the boson in the loop.  Also, the self-energy
in the one-boson sector is different from the 
self-energy correction in bare-fermion sector.
In the LFCC equations, the self-energy corrections
are the same everywhere they appear and are not
spectator dependent.  The price paid for this gain
is the nonlinear nature of the equation for $t_1$.

\begin{figure}
\begin{center}
\begin{tabular}{c}
\includegraphics[height=3cm]{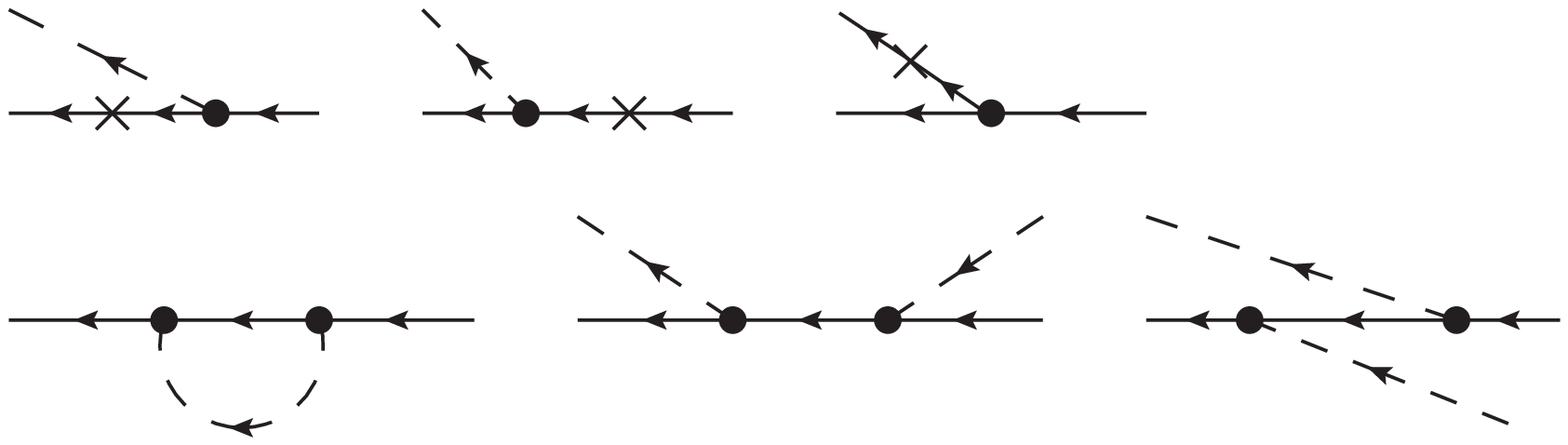} \\ (a) \\
\includegraphics[height=1.5cm]{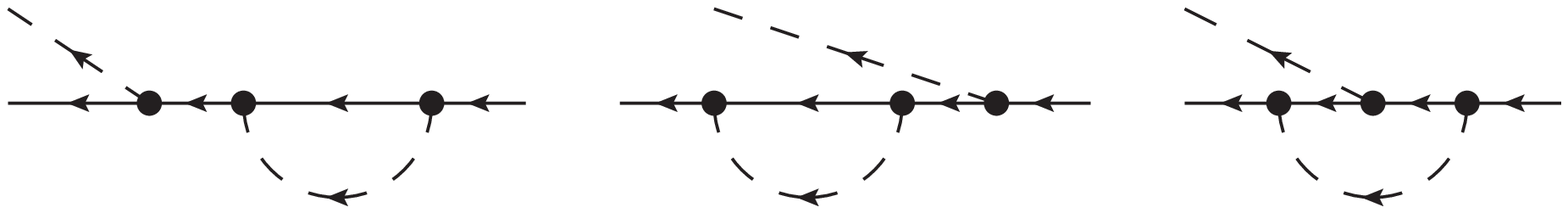} \\ (b)
\end{tabular}
\end{center}
\caption{\label{fig:commutators}
The commutators (a) $[\Pminus,T]$ and (b) $[[\Pminus,T],T]$
used to define the Baker--Hausdorff expansion of
the LFCC effective Hamiltonian $\ob{\Pminus}$.  Only terms
that connect the lowest Fock sectors, or contribute to such
terms, are listed.  In particular, the last two diagrams of
(a) contribute to the commutator in (b).}
\end{figure}

To derive the LFCC equations, we must first construct
the effective Hamiltonian.  This
is done by computing the commutators of the Baker--Hausdorff
expansion for $\ob{\Pminus}$.  The necessary commutators
are represented in Fig.~\ref{fig:commutators}.  When
added to $\Pminus$, they provide all that is needed
to define the LFCC eigenvalue problem and auxiliary
equations for the bare-fermion valence state with 
the chosen truncation of $T$.  Notice that all three
of the diagrams analogous to those for the Ward
identity in QED are included.

One can show that the truncation for $T$ provides an
exact solution for the model~\cite{LFCC}, so that this
lowest-order LFCC approximation is actually exact.
Solution of the left-hand eigenproblem, for $\langle\widetilde\psi|$,
then permits calculation of the Dirac form factor from Eq.~(\ref{eq:DiracFF}).
Details are given in \cite{LFCC}.

\section{Light-front holographic QCD} \label{sec:holographicQCD}

The LFCC valence-sector eigenvalue problem for mesons and baryons 
in QCD can be approximated by models based on light-front holography~\cite{holographicQCD}.
A factorized meson wave function in the valence ($q\bar{q}$) sector
\be
\psi=e^{iL\varphi}X(x)\phi(\zeta)/\sqrt{2\pi\zeta}
\ee
is subject to an effective potential $\widetilde U$ that conserves $L_z$
\be
\left[ \frac{\mu_1^2}{x}+\frac{\mu_2^2}{1-x}-\frac{\partial^2}{\partial\zeta^2}
-\frac{1-4L^2}{4\zeta^2}+\widetilde U\right]X(x)\phi(\zeta)=M^2X(x)\phi(\zeta).
\ee
For zero-mass quarks, the longitudinal wave function $X$ decouples,
and the transverse wave function satisfies
\be \label{eq:transverse}
\left[ -\frac{d^2}{d\zeta^2}
-\frac{1-4L^2}{4\zeta^2}+ U(\zeta)\right]\phi(\zeta)=M^2\phi(\zeta),
\ee
with $U$ determined by an AdS$_5$ correspondence.

The softwall model for massless quarks~\cite{softwall}
yields an oscillator potential $U(\zeta)=\kappa^4\zeta^2+2\kappa^2(J-1)$
and a simple spectrum.  The masses $M^2=4\kappa^2\left(n+(J+L)/2\right)$
have a linear Regge trajectory and are a good fit for light 
mesons~\cite{Gershtein,Vega,Ebert,Branz,Kelley}.
The transverse wave functions are 2D oscillator functions.
The longitudinal wave function $X$ is constrained by a form-factor 
duality~\cite{formfactormatch}, between the Fock-space construction
\be
F(q^2)=\int \frac{dx\,|X(x)|^2}{x(1-x)} \int 2\pi \zeta d\zeta 
     J_0\left(\zeta q_\perp\sqrt{x/(1-x)}\right)|\phi(\zeta)|^2
\ee
and the form computed in AdS$_5$
\be
F(q^2)=\int dx \int 2\pi \zeta d\zeta 
     J_0\left(\zeta q_\perp\sqrt{x/(1-x)}\right)|\phi(\zeta)|^2.
\ee
Thus $X(x)=\sqrt{x(1-x)}$, when the quarks are massless.

For massive quarks, there is the {\em ansatz} by 
Brodsky and De T\'eramond~\cite{ansatz}, to
replace $k_\perp^2/(x(1-x))$ with
$k_\perp^2/x(1-x)+\mu_1^2/x+\mu_2^2/(1-x)$
in the transverse harmonic oscillator eigenfunctions,
with $\mu_i$ as current-quark masses.  This yields
a longitudinal wave function of the form
\be \label{eq:ansatz}
X_{\rm BdT}(x)=N_{\rm BdT}\sqrt{x(1-x)}e^{-(\mu_1^2/x+\mu_2^2/(1-x))/2\kappa^2},
\ee

Instead of this {\em ansatz}, we use a longitudinal equation for $X$,
with an effective potential from the `t~Hooft model~\cite{tHooft}
\be \label{eq:longitudinaleqn}
\left[\frac{m_1^2}{x}+\frac{m_2^2}{1-x}\right]X(x)
  +\frac{g^2}{\pi}{\cal P}\int dy \frac{X(x)-X(y)}{(x-y)^2}-C X(x)=M_\parallel^2 X(x),
\ee
where the $m_i$ are constituent masses.  The 't~Hooft model, which is based on
large-$N$ two-dimensional QCD, incorporates longitudinal confinement in a manner
consistent with four-dimensional QCD.
The solution of this equation, $X(x)$, is known~\cite{tHooft,Bergknoff}
to be well approximated by $x^{\beta_1}(1-x)^{\beta_2}$, with the $\beta_i$
subject to the constraints
$m_i^2 \pi/g^2-1+\pi\beta_i \cot\pi\beta_i=0$.
For consistency, we should have $\beta_1=\beta_2=1/2$ in
the zero-current-mass limit.  This fixes the coupling 
to be $g^2/\pi=m_u^2=m_d^2$.

The longitudinal equation is relatively easy to solve numerically~\cite{Bergknoff,MaHiller,MoPerry}.
We expand the solution as $X(x)=\sum_n c_n f_n(x)$ with respect to
basis functions constructed from Jacobi polynomials~\cite{MoPerry}
\be  \label{eq:basis}
f_n(x)=N_n x^{\beta_1}(1-x)^{\beta_2}P_n^{(2\beta_2,2\beta_1)}(2x-1).
\ee
The $n=0$ term represents 90\% or more of the probability.
The matrix representation of the longitudinal equation, for $M_\parallel=0$,
is then
\be
\left( \frac{m_1^2}{m_u^2}A_1+\frac{m_2^2}{m_u^2}A_2+B\right)\vec c=\xi\vec c, 
\ee
with $\xi\equiv C/m_u^2$ and
\bea
(A_1)_{nm}&=&\int_0^1 \frac{dx}{x} f_n(x) f_m(x),\;\;
(A_2)_{nm}=\int_0^1 \frac{dx}{1-x} f_n(x) f_m(x), \\
B_{nm}&=&\frac12\int_0^1 dx \int_0^1 dy 
   \frac{f_n(x)-f_n(y)}{x-y}\frac{f_m(x)-f_m(y)}{x-y}.
\eea
The solution of the matrix problem yields the coefficients
for the basis-function expansion.

\begin{figure}
\vspace{0.2in}
\begin{center}
\begin{tabular}{cc}
\includegraphics[width=7cm]{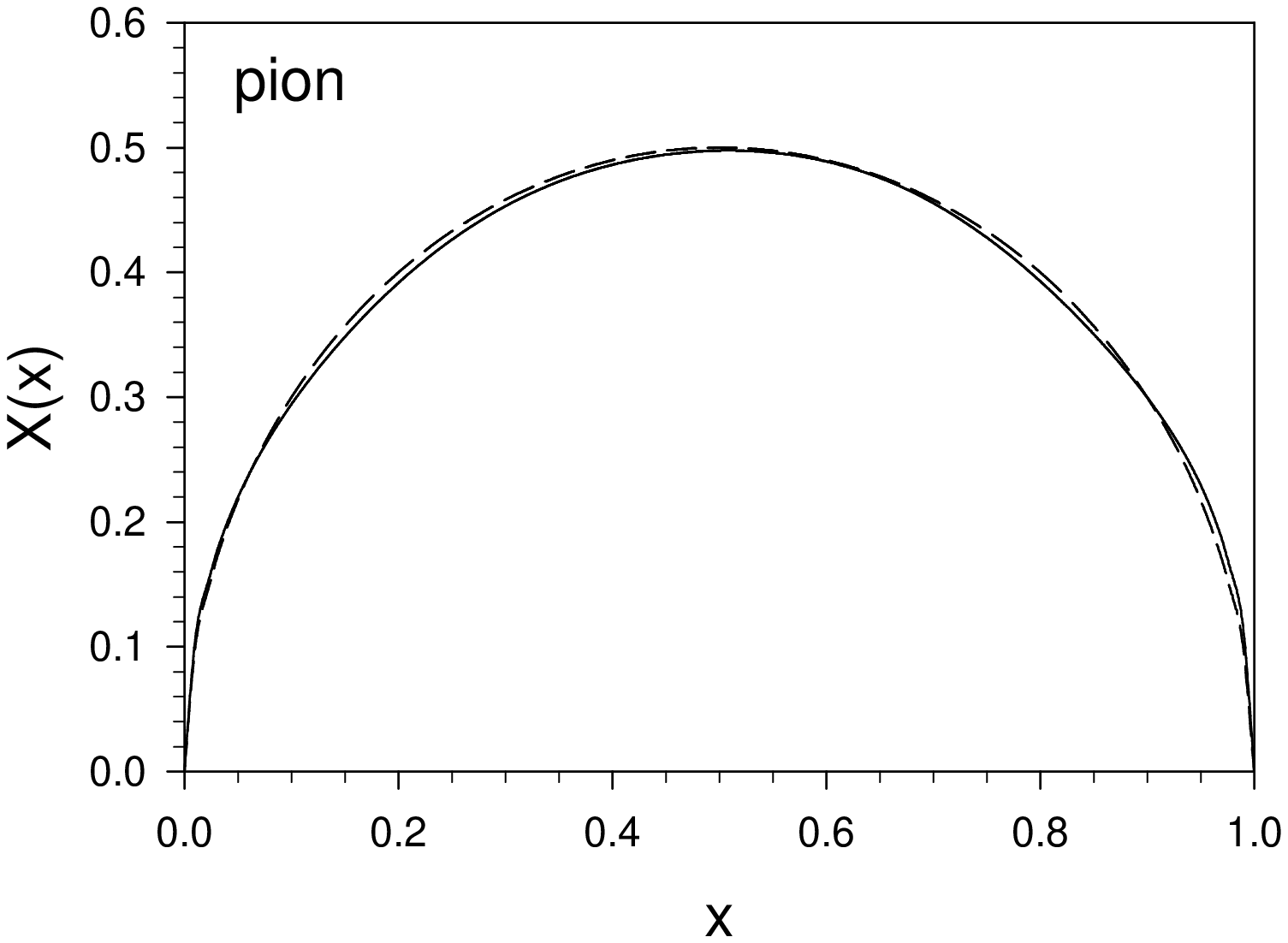} &
\includegraphics[width=7cm]{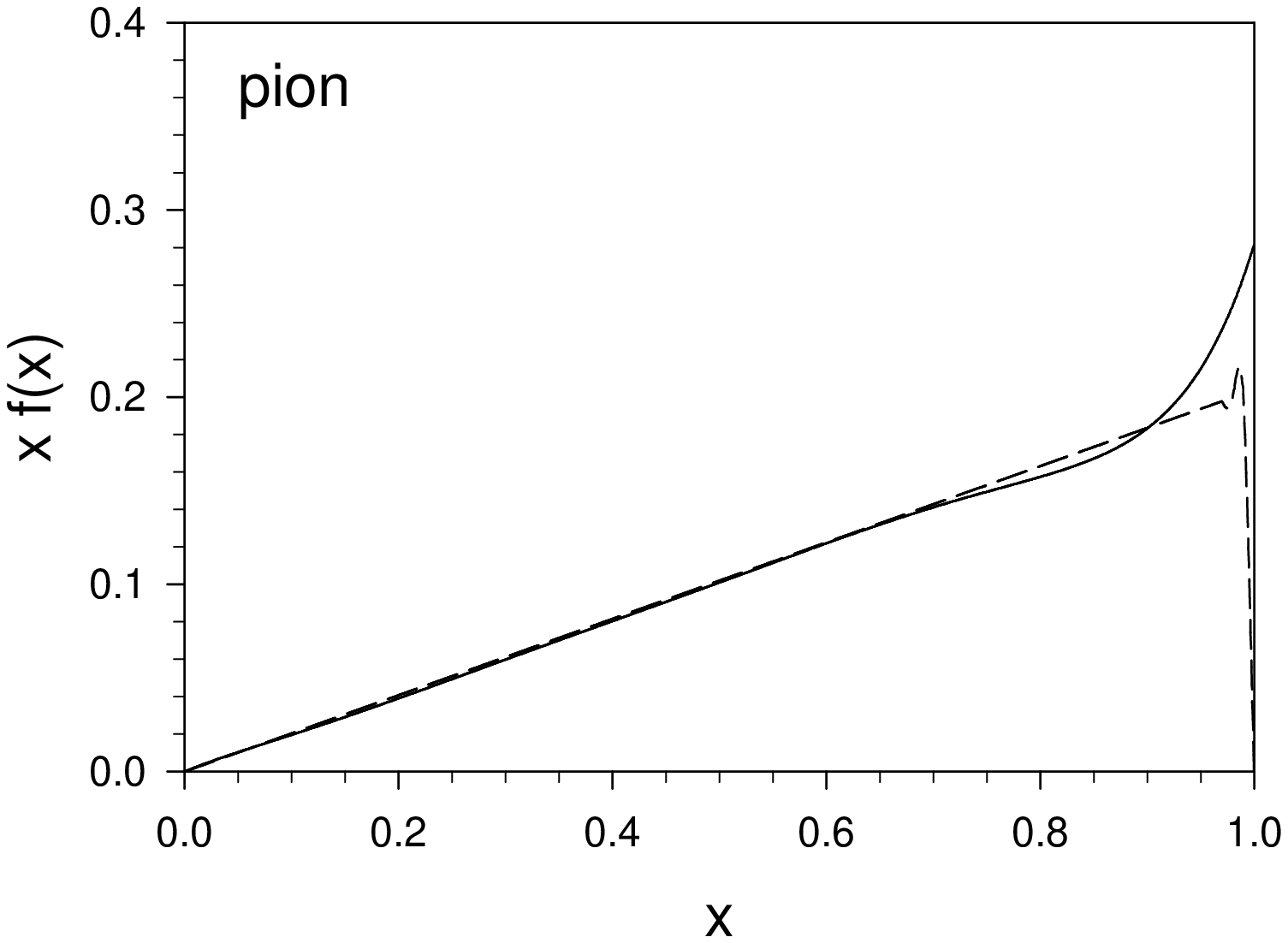} \\
(a) & (b) 
\end{tabular}
\end{center}
\caption{\label{fig:pion}
Longitudinal wave function (a) and parton distribution function (b)
for the pion.
The solid lines are
wave functions from our model; the dashed lines show the {\em ansatz}
by Brodsky and De~T\'eramond~\protect\cite{ansatz}.
}
\end{figure}

\begin{figure}
\vspace{0.2in}
\begin{center}
\begin{tabular}{cc}
\includegraphics[width=7cm]{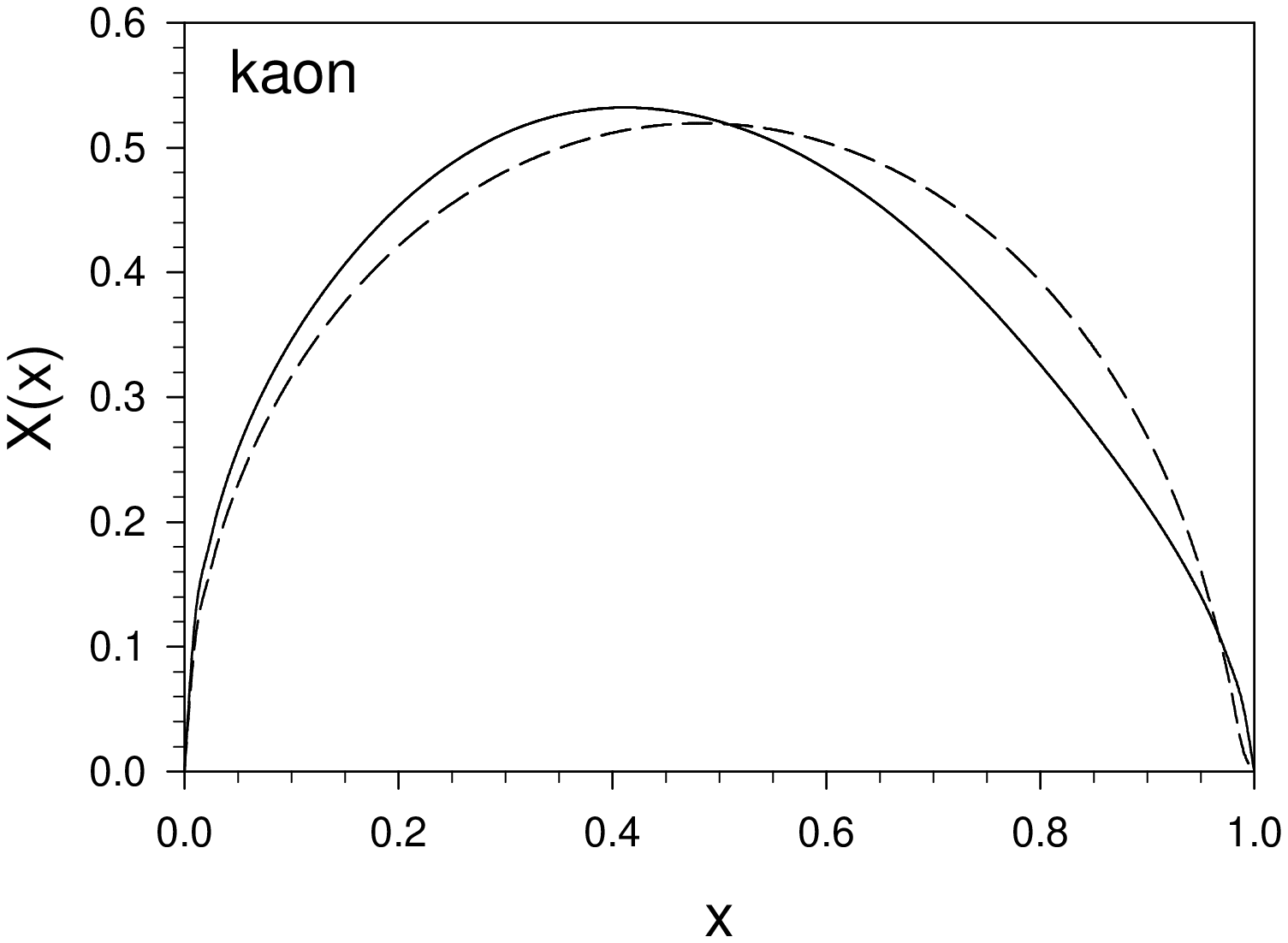} &
\includegraphics[width=7cm]{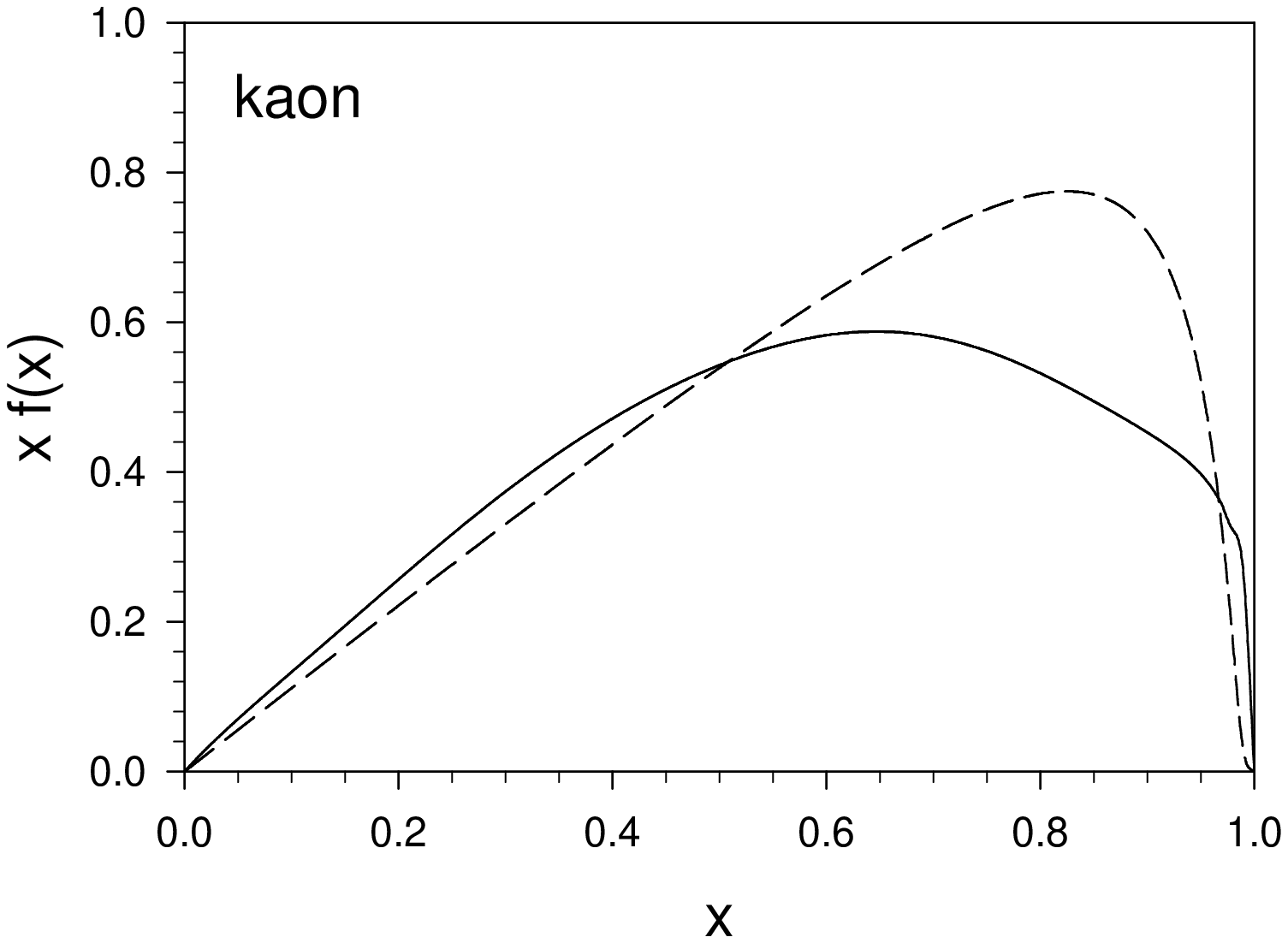} \\
(a) & (b) 
\end{tabular}
\end{center}
\caption{\label{fig:kaon}
Same as Fig.~\ref{fig:pion}, but for the kaon.}
\end{figure}

\begin{figure}
\vspace{0.2in}
\begin{center}
\begin{tabular}{cc}
\includegraphics[width=7cm]{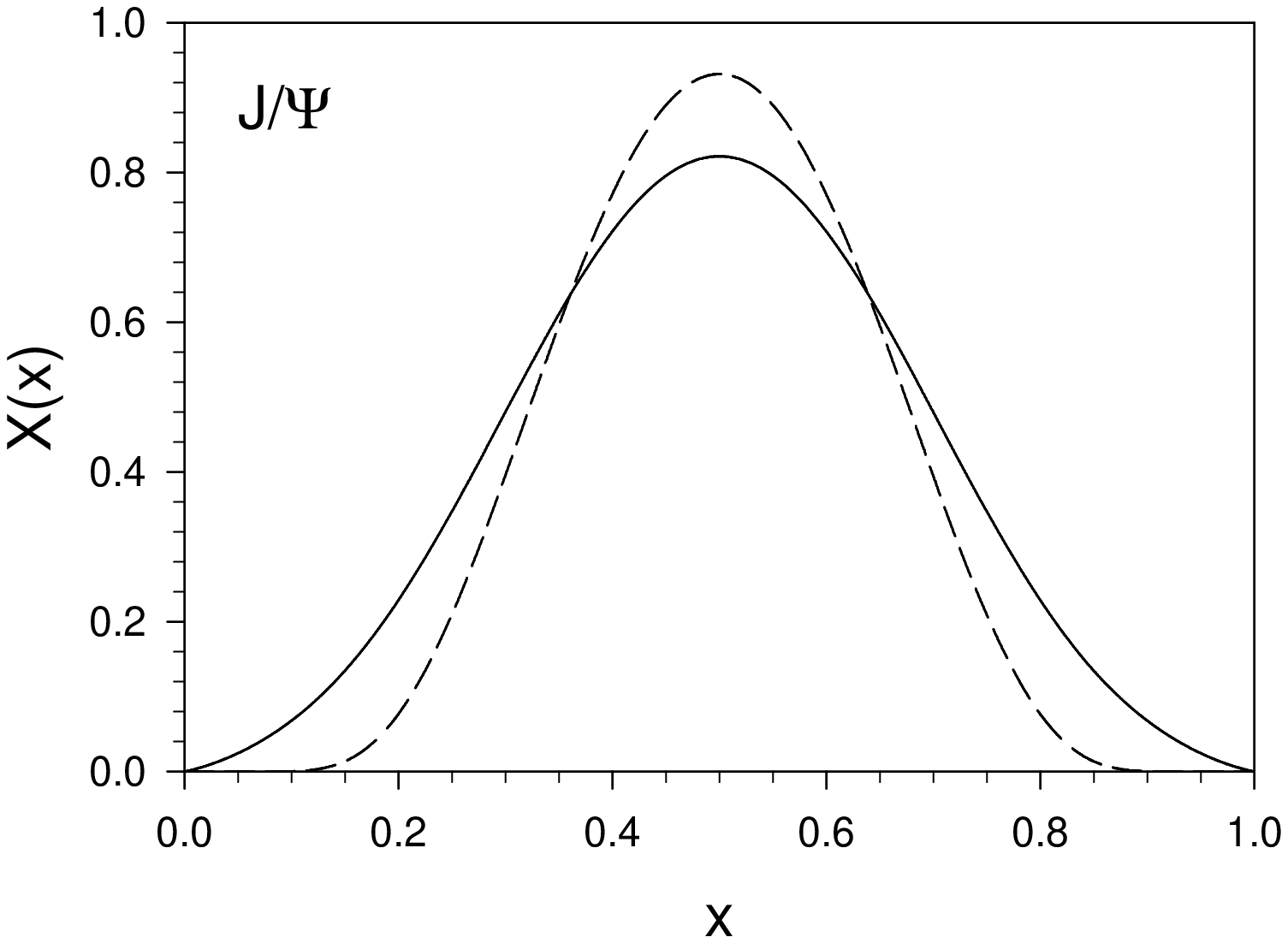} &
\includegraphics[width=7cm]{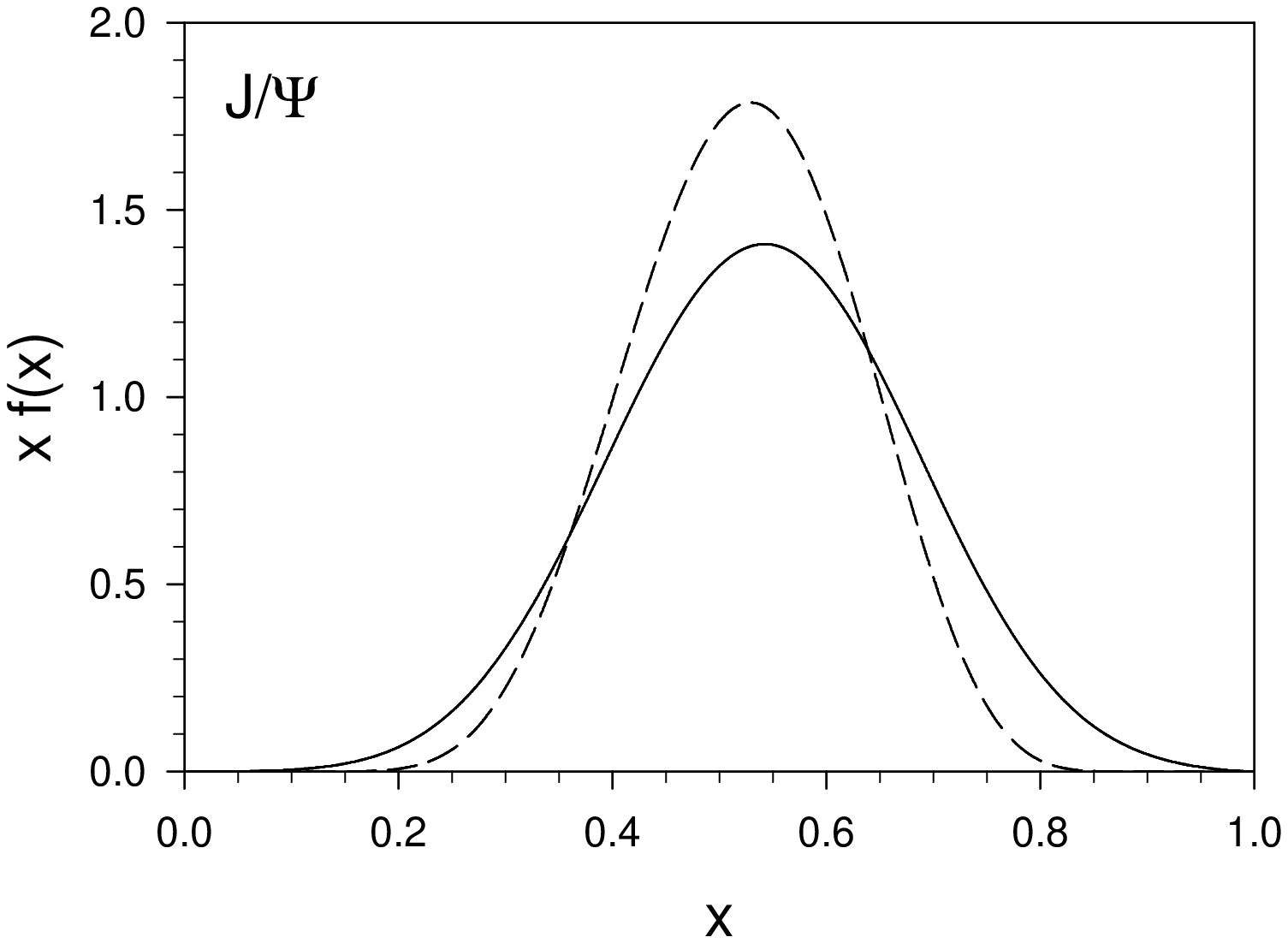} \\
(a) & (b) 
\end{tabular}
\end{center}
\caption{\label{fig:JPsi}
Same as Fig.~\ref{fig:pion}, but for the J/$\Psi$.}
\end{figure}

The wave function can the be used to compute the
decay constant~\cite{decayconstant}
\be
f_M=2\sqrt{6}\int_0^1 dx \int_0^\infty \frac{dk_\perp^2}{16\pi^2} \psi(x,k_\perp)
\ee
and the parton distribution~\cite{Vega,pdf}
\be
f(x)=P_{q\bar{q}}\frac{X^2(x)}{x(1-x)},
\ee
where $P_{q\bar{q}}$ is the probability of the $q\bar{q}$
Fock component in the meson.
The chosen parameter values and the resulting decay constants
are listed in Table~\ref{tab:params} for the pion, kaon,
and J/$\Psi$.  The wave functions and parton distributions
are very similar to those of the {\em ansatz} (\ref{eq:ansatz}), 
except for the J/$\Psi$, as can be seen in Figs.~\ref{fig:pion},
\ref{fig:kaon}, and \ref{fig:JPsi}.

\begin{table}
\caption{\label{tab:params}Parameters and decay constants, compared with the {\em ansatz} of Brodsky and De T\'eramond~\protect\cite{ansatz}.
All dimensionful parameters are in units of MeV. Parameter and experimental values
are from Ref.~\protect\cite{Vega}
and the Particle Data Group~\protect\cite{PDG}.
}
\begin{center}
\begin{tabular}{cccccccccc}
\hline \hline
 & \multicolumn{2}{c}{model} &  \multicolumn{2}{c}{\em ansatz} &
  &   &  \multicolumn{3}{c}{decay constant} \\
meson & $m_1$ & $m_2$ & $\mu_1$ & $\mu_2$ & $P_{q\bar{q}}$ & $\kappa$ & model & {\em ansatz} & exper. \\
\hline
pion & 330 & 330 & 4 & 4 & 0.204 & 951 & 131 & 132 & 130 \\
kaon & 330 & 500 & 4 & 101 & 1 & 524 & 160 & 162 & 156 \\
J/$\Psi$ & 1500 & 1500 & 1270 & 
  1270 & 1 & 894 & 267 & 238 & 278 \\
\hline \hline
\end{tabular}
\end{center}
\end{table}

\section{Summary} \label{sec:summary}

In order to avoid truncation of Fock space, the
LFCC method~\cite{LFCC} generates all the higher Fock-state
wave functions from the lower wave functions,
based on the solution of nonlinear equations
for vertex-like functions.  This eliminates
sector dependence and spectator dependence from
the terms in the effective Hamiltonian.  The
truncation of the nonlinear equations can be
relaxed systematically, to provide ever more
sophisticated approximations for the higher
wave functions.

The LFCC method divides the hadronic eigenproblem
into an effective eigenproblem in the valence sector
and auxiliary equations that define the effective
Hamiltonian.  Light-front holography~\cite{holographicQCD} then provides
a model for the valence sector.  This model can
be augmented to include quark masses and a 
dynamical equation for the longitudinal wave function~\cite{LongWF}
that is consistent with the Brodsky-de T\'eramond
{\em ansatz}~\cite{ansatz}.  The numerical solution of the longitudinal
equation includes a choice of basis functions that could
be useful beyond just the holographic approximation to QCD.

\acknowledgments
This work was done in collaboration with S.S. Chabysheva
and supported in part by the US Department of Energy and 
the Minnesota Supercomputing Institute.

\end{document}